\documentclass[aps,pra,twocolumn,showpacs,groupedaddress,bibnotes]{revtex4-1}

\usepackage{graphicx}
\usepackage{bm}
\usepackage{amsmath}
\usepackage{textcomp}

\graphicspath{{pict/}{}}
\newcommand\pictc[5]{\begin{figure}
            \centerline{\vspace{0mm}
\includegraphics[width=#1\columnwidth,height=0.7\textheight,keepaspectratio]{#3}}
            \protect\caption{\protect\label{fig:#4} #5}\vspace{-0mm}
                    \end{figure}            }

\newcommand\pict[4][1.0]{\pictc{#1}{h}{#2}{#3}{#4}}

\newcommand\rpict[1]{\ref{fig:#1}}

\newcommand\leqt[1]{\protect\label{eq:#1}}
\newcommand\reqtn[1]{\ref{eq:#1}}
\newcommand\reqt[1]{(\reqtn{#1})}

\newcommand\lsect[1]{\protect\label{sect:#1}}
\newcommand\rsect[1]{\ref{sect:#1}}

\newcounter{Fig}

\newcommand{\ee}{{\rm e}}

\begin{document}
\begin{sloppy}

\title{Light Bullets in Nonlinear Periodically Curved Waveguide Arrays}

\author{Michal~Matuszewski$^{1}$, Ivan~L.~Garanovich$^{1}$, and Andrey~A.~Sukhorukov$^{1,2}$}

\affiliation{$^1$Nonlinear Physics Centre and $^2$Centre for Ultra-high bandwidth Devices for Optical Systems (CUDOS),
Research School of Physics and Engineering,\\ Australian National University, Canberra, ACT 0200, Australia
}

\begin{abstract}
We predict that stable mobile spatio-temporal solitons can exist in arrays of periodically curved optical waveguides.
We find  two-dimensional light bullets in one-dimensional arrays with harmonic waveguide bending and three-dimensional bullets in square lattices with helical waveguide bending using variational formalism.
Stability of the light bullet solutions is confirmed by the direct numerical simulations which show that the light bullets can freely move across the curved arrays. This mobility property is a distinguishing characteristic compared to previously considered discrete light bullets which were trapped to a specific lattice site. These results suggest new possibilities for flexible spatio-temporal manipulation of optical pulses in photonic lattices.
\end{abstract}

\pacs{42.65.Tg, 42.82.Et}

\maketitle

\section{Introduction} \lsect{Introduction}

Discrete solitons, the self-localized states which can exist in both one- and two-dimensional nonlinear photonic lattices, have attracted a lot of attention over the recent years~\cite{Lederer:2008-1:PRP}.
The interplay between the underlying periodicity and nonlinear self-action enables a whole lot
of new opportunities for all-optical control, which are not possible in homogeneous media~\cite{Lederer:2008-1:PRP}.
Of particular interest are pulses which are localized
both in space and in time as a result of balance between the self-focusing nonlinearity, diffraction, and anomalous dispersion.
Such spatio-temporal optical solitons are called light bullets~\cite{Silberberg:1990-1282:OL}.
Light bullets are inherently multi-dimensional, and consequently they are always unstable in Kerr media~\cite{Silberberg:1990-1282:OL}.
Discreteness in photonic lattices can provide an additional stabilization mechanism, and indeed stable continuous-discrete solitons were
found in nonlinear waveguide arrays~\cite{Laedke:1995-5549:PRE, Baizakov:2004-53613:PRA}.
However, these discrete light bullets are strongly trapped in a few waveguide channels, and hence their mobility
is practically reduced to the longitudinal dimension, whereas they cannot move across the photonic lattice.

Recently, it was suggested that arrays of periodically curved optical waveguides offer unique opportunities for the control of both the strength and frequency dispersion of the waveguide coupling~\cite{Eisenberg:2000-1863:PRL, Longhi:2006-243901:PRL, Szameit:2009-271:NAPH, Iyer:2007-3212:OE}. Propagation of light beams in various types of the nonlinear lattices with modified diffraction has been analyzed~\cite{Konotop:1993-563:PRE, Cai:1995-1186:PRL, Pertsch:2003-744:CHA, Longhi:2005-2137:OL, Staliunas:2007-11604:PRA, Szameit:2009-153901:PRL}. It was demonstrated that light beams can become localized in the periodically curved nonlinear waveguide arrays in the form of diffraction-managed solitons~\cite{Ablowitz:2001-254102:PRL, Garanovich:2007-9547:OE, Szameit:2008-31801:PRA}. These diffraction-managed spatial solitons are reminiscent of dispersion-managed temporal solitons~\cite{Gabitov:1996-327:OL, Stratmann:2005-143902:PRL, Malomed:2006:SolitonManagement}.
However, nonlinear propagation and localization of light { pulses} in diffraction managed photonic lattices have not been analyzed yet.

In this work, we predict, for the first time to the best of our knowledge, that {\em stable broad spatio-temporal solitons can exist in arrays of periodically curved optical waveguides,} and such discrete light bullets {\em are  mobile and can move across the array.} We find approximate light bullet solutions using the variational technique, where we assume a Gaussian shape of the spatial-temporal soliton as shown in Fig.~\rpict{var2d}(b). The results of the variational analysis indicate that by optimizing the waveguide bending amplitude and period, one can achieve stable and mobile two- and three-dimensional light bullets.
We emphasize that we find stable light bullets which spatial width is much larger than the
lattice period. Hence, the light bullets would not be trapped in a single waveguide, retaining mobility in the transverse dimension. This conclusion is confirmed by direct numerical simulations. We also demonstrate that a similar approach can be used to achieve stabilization of multidimensional matter-wave solitons in Bose-Einstein condensates.

\section{Two-dimensional light bullets} \lsect{2d}
\subsection{Discrete-continuous model equation} \lsect{2d-discrete}

First, we consider propagation of light pulses in a one-dimensional array of coupled optical waveguides, where waveguide axes are periodically curved in the longitudinal propagation direction~$z$,
see the sketch in Fig.~\rpict{var2d}(a).
We model the light bullet dynamics with the modified discrete-continuous nonlinear Schrodinger propagation equation which takes into the periodic waveguide bending~\cite{Longhi:2005-2137:OL, Garanovich:2007-9737:OE} and temporal dispersion~\cite{Peschel:2002-2637:JOSB, Droulias:2005-1827:OE, Lahini:2007-23901:PRL, Heinrich:2009-113903:PRL}:
\begin{align}  \leqt{PE}
     i \frac{\partial a_n}{\partial z}&
    + C \left(\ee^{-i p(z)} a_{n+1} + \ee^{i p(z)} a_{n-1}\right) +\nonumber\\
    & + \frac{D}{2} \frac{\partial^2 a_n}{\partial t^2}
    + \gamma |a_n|^2 a_n
    = 0 \,.
\end{align}
Here $n$ is the waveguide number, $a_n$ are the complex amplitudes of individual waveguide modes, $z$ is the propagation distance, $t$ is time, $C$ is the coupling coefficient between the modes of straight neighboring waveguides, $D$ is the second-order dispersion coefficient, $\gamma$ is the nonlinear coefficient, and $p(z)$ is the phase coefficient related to the waveguide bending~\cite{Garanovich:2007-9737:OE}.
We consider all variables in Eq.~\reqt{PE} to be dimensionless, applying the following normalization:
\begin{equation}  \leqt{PEnorm}
    \begin{array}{l} {\displaystyle
        a_n =  A_n / \sqrt{I_s} , \,
        z = z_d / z_s, \,
        t = t_d / t_s, \,
    } \\*[9pt] {\displaystyle
        C = C_d z_s, \,
        D = D_d z_s t_s^{-2}, \,
        \gamma = \gamma_d I_s z_s, \,
    } \end{array}
\end{equation}
where $A_n$ and variables with the subscript $d$ denote the values in physical dimensions, and subscript $s$ denotes the scaling coefficients. With these notations, the coupling phase coefficient is defined as
\begin{equation}  \leqt{PEphase}
    p(z) = \frac{2 \pi n_0 d x_s^2}{\lambda_d z_s} x^{\prime}_0
\end{equation}
where $x_0(z)$ is the waveguide bending profile normalized to $x_s$, prime stands for the derivative, $d_x = d_d / x_s$ is the normalized transverse separation between the neighboring waveguides, $\lambda_d$ is the laser wavelength in vacuum, and $n_0$ is the average medium refractive index. We note that the values of the scaling coefficients $x_s$, $z_s$, $t_s$, and $I_s$ can be chosen arbitrarily. Accordingly, the results obtained for a specific set of dimensionless parameters can be related to a variety of physical configurations.

\subsection{Continuous approximation for mobile solitons} \lsect{2d-continuous}

As outlined in the introduction, our aim is to identify the conditions for the existence of stable and mobile solutions in the form of light bullets, which remain localized in space and time as they propagate along the waveguide array. It was found that
the solitons which are strongly localized in the spatial domain tend to become trapped at a particular lattice site~\cite{Kivshar:1993-3077:PRE, Morandotti:1999-2726:PRL}. Only when the soliton width extends over five or more waveguides, the soliton becomes mobile~\cite{Aceves:1996-1172:PRE}. However, such broad spatio-temporal solitons were found to be unstable~\cite{Laedke:1995-5549:PRE, Baizakov:2004-53613:PRA}.

In order to study the key effects of periodic waveguide bending on the stability of broad spatio-temporal light bullets, we employ the continuous approximation. Specifically, we consider wavepackets that are slowly varying in space, and describe their profiles with a continuous function,
$a_n(t,z) = \tilde{u}(x=n d,t,z)$.
We expand $\tilde{u}$ in Taylor series in $x$ and substitute in Eq.~\reqt{PE},
\begin{align}
    i \frac{\partial \tilde{u}}{\partial z}&
    + 2 C\left[\left(\tilde{u}
    + \frac{d^2}{2} \frac{\partial^2 \tilde{u}}{\partial x^2} \right)\cos p(z)
    - i d \frac{\partial \tilde{u}}{\partial x} \sin p(z) \right] +\nonumber\\
    & + \frac{D}{2} \frac{\partial^2 \tilde{u}}{\partial t^2} + \gamma |\tilde{u}|^2 \tilde{u} =0\,.
\end{align}
Then, we make a transformation
\begin{equation} \leqt{uTransfort}
    \tilde{u}(x,t,z) = u(x-x_1(z),z) \exp\left[ 2 i C \int_0^{z} \cos p(z_1) dz_1  \right] ,
\end{equation}
where
$$x_1(z) = 2 C d \int_0^z \sin p(z_1) dz_1,$$
and finally obtain
\begin{align} \leqt{NLSuGeneral}
    i \frac{\partial u}{\partial z}
    + C d_x^2 \cos p(z) \frac{\partial^2 u}{\partial x^2}
    + \frac{D}{2} \frac{\partial^2 u}{\partial t^2}
    + \gamma |u|^2 u
    = 0\,.
\end{align}

We further consider the harmonic waveguide bending profile,
\begin{equation} \leqt{x0Harmonic}
    x_0(z) = x_{0m} \cos ( 2 \pi z / L ),
\end{equation}
where $x_{0m}$ is the bending amplitude (normalized to $x_s$) and $L = L_{\rm_d} / z_s$ is the normalized bending period.
To reduce the number of independent coefficients in Eq.~\reqt{NLSuGeneral}, we choose the scaling coefficients as follows:
\begin{equation} \leqt{scaling}
    z_s = \frac{L_d}{2 \pi},\,
    t_s = \sqrt{|D_d| z_s},\,
    x_s = d_d \sqrt{2 C_d z_s},\,
    I_s = \frac{1}{\gamma_d z_s}.
\end{equation}
Then, the normalized coefficients are
\begin{equation} \leqt{scalingCoeff}
    \begin{array}{l} {\displaystyle
          L=2\pi,\,
          d_x = \frac{1}{2 C^{2}},
    } \\*[9pt] {\displaystyle
          D=\sigma_D = {\rm sign}(D_d) = \pm 1,\,
    } \\*[9pt] {\displaystyle
          \gamma = \sigma_\gamma = {\rm sign}(\gamma_d) = \pm 1,\,
    } \end{array}
\end{equation}
and Eq.~\reqt{NLSuGeneral} can be written as
\begin{align} \leqt{PE_red}
    i \frac{\partial u}{\partial z}
    + \frac{\lambda(z)}{2}  \frac{\partial^2 u}{\partial x^2}
    + \frac{\sigma_D}{2} \frac{\partial^2 u}{\partial t^2}
    + \sigma_\gamma |u|^2 u =0 \,,
\end{align}
where
\begin{equation} \leqt{lambda}
    \lambda(z) = \cos (B \sin z),\,
    B = - x_{0m} \frac{4 \pi^2 n_0 d_x x_s^2}{\lambda_d z_s}.
\end{equation}
In self-focusing media we have $\sigma_\gamma =1$, whereas $\sigma_\gamma = -1$ in case of self-defocusing nonlinearity. In order to achieve soliton localization in temporal dimension we require that the temporal dispersion is either anomalous in self-focusing media~\cite{Silberberg:1990-1282:OL}, or dispersion is normal if the nonlinearity is self-defocusing, i.e. $\sigma_D \sigma_\gamma > 0$.

We note that there is just one parameter, the coupling coefficient $C$, which defines the mapping of solutions based on the continuous Eq.~\reqt{PE_red} to the original discrete-continuous Eq.~\reqt{PE}. Continuous solutions may serve as a good approximation provided that their spatial width ($W$) is much larger than the distance between the neighboring waveguides,
\begin{equation} \leqt{widthCondition}
  W \gg d_x = \frac{1}{2 C^{2}} = \frac{2 \pi^2}{C_d^2 L_d^2} = \frac{2 L_c^2}{L_d^2} ,
\end{equation}
where we introduce the notation $L_c = \pi / C_d$ to define the physical coupling length between two straight waveguides~\cite{Yariv:2006:Photonics}. It follows that Eq.~\reqt{PE_red} can be used as an approximate model for Eq.~\reqt{PE} provided that the normalized coupling coefficient ($C$), or equivalently the ratio of the bending period to the coupling length, are sufficiently large.

\pict[1.0]{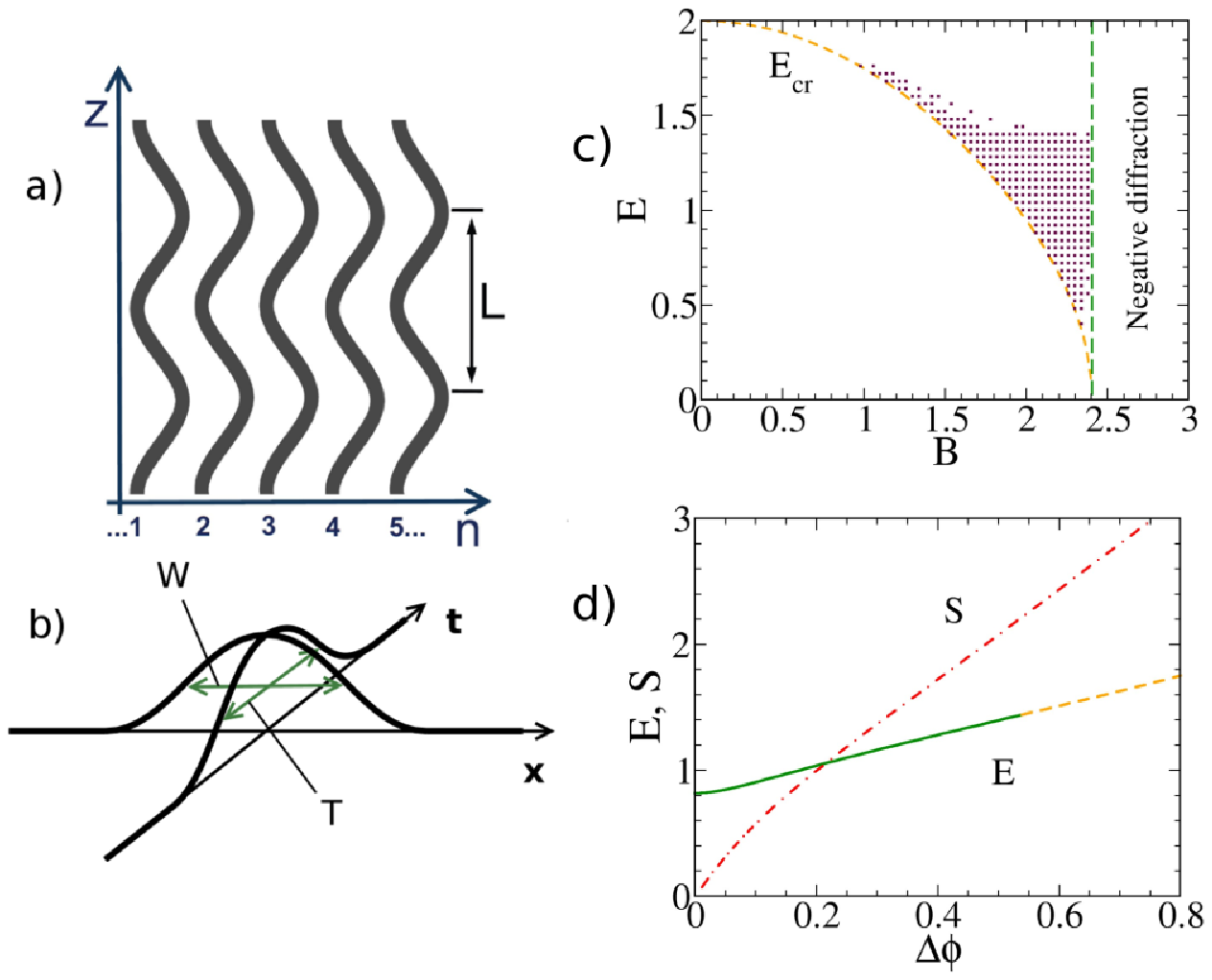}{var2d}{
(a) Schematic of a periodically curved one-dimensional waveguide array. (b) Profile of a spatiotemporal light bullet with
Gaussian ansatz.
(c) Stability diagram in the two-dimensional case. The dashed lines indicate the value of the critical energy $E_{\rm cr}$ and the
boundary between positive and negative average diffraction regions $\langle\lambda\rangle$.
(d)~The dependence of $E$ vs. $\Delta \phi$ (phase increment after one period) for $B=2.1$.
Solid line corresponds to stable solutions, dashed line to unstable solutions.
The dash-dotted line indicates the diffraction map strength $S$.
}

\subsection{Variational approach} \lsect{2d-variational}

We now look for approximate analytical solutions of the Eq.~\reqt{PE_red} using the variational method~\cite{Malomed:2002-71:ProgressOptics}.
We consider the Gaussian Ansatz for the light bullet spatiotemporal profile [see Fig.~\rpict{var2d}(b)],
\begin{eqnarray}
    u & = & A(z)
     \exp \left\{ \mathrm{i}\phi (z)-\frac{1}{2}\left[\frac{x^{2}}{W^{2}(z)}+\frac{t
            ^{2}}{T^{2}(z)}\right] \right. +
        \nonumber \\
    &&+\left. \frac{\mathrm{i}}{2}\left[ b(z)\,x^{2} +\beta (z)\,t
        ^{2}\right] \right\} ,  \leqt{ansatz}
\end{eqnarray}
where $A$ and $\phi $ are the amplitude and phase of the soliton,
$T$ and $W$ are its temporal and transverse spatial widths, and $\beta $
and $b$ are the temporal and spatial chirps. The Eq.~\reqt{PE_red}
can be derived from the Lagrangian
\begin{eqnarray}
    {\cal L} &=&\frac{1}{2}\int\!\!\int\,\mathrm{d}x\mathrm{d}t \left[
    \mathrm{i}\left( u_{z}u^{\ast }-u_{z}^{\ast }u\right) -\lambda\left\vert
    u_{x}\right\vert ^{2}-\left\vert u_{t
    }\right\vert ^{2}\right.  \nonumber\\
    &&\left. +|u|^{4}\right] . \leqt{Lagrangian}
\end{eqnarray}
We substitute the ansatz Eq.~\reqt{ansatz} into Eq.~\reqt{Lagrangian} and
integrate over spatial coordinates to obtain the reduced  Lagrangian
\begin{eqnarray}
    (4/\pi){\cal L}_{\mathrm{eff}} &=&A^{2}WT\left[ 4\phi ^{\prime }-b^{\prime
    }W^{2}-\beta ^{\prime }T^{2}-W^{-2}-DT^{-2}\right.
    \nonumber\\
    &&\left. -\lambda(z)b^{2}W^{2}
    -\beta ^{2}T^{2}+A^{2}/\sqrt{2}\right] \,.
\end{eqnarray}
which variation with respect to $\phi$
yields the energy conservation $dE/dz=0$ with the constant $E$ being proportional to the pulse energy
\begin{equation} \leqt{E}
    E\equiv A^{2}WT = \frac{1}{\pi} \int\!\!\!\int |u|^2 dx dt.
\end{equation}
The conservation of $E$ is used to eliminate $A$ from the remaining variational equations
\begin{eqnarray}
    W^{\prime }&=&\lambda b W,   \leqt{variat1} \\
    b^{\prime } &=&\frac{\lambda}{W^{4}}-\lambda b^2 -\frac{E}{2W^{3}T},  \leqt{VA_2D} \\
    T^{\prime \prime } &=&\frac{1}{T^{3}}-\frac{E}{2WT^{2}}, \leqt{variat3}
\end{eqnarray}
where the equations for $\beta^{\prime }$ and $T^{\prime }$ were used to obtain the single second order equation for $T^{\prime \prime }$.

The systematic results obtained using  the variational model Eq.~\reqt{VA_2D} are shown in Fig.~\rpict{var2d}(c). 
The dotted area in Fig.~\rpict{var2d}(c) corresponds to the values of parameters $E$ and $B$ for which stable light bullet solutions exist which oscillate with the period of the waveguide modulation. The solutions of the variational equations were obtained using Newton relaxation method. The two boundaries of the stability region can be found analytically, they are indicated in Fig.~\rpict{var2d}(c) by the two dashed lines . The first vertical line on the right corresponds to the value of waveguide bending $B$ for which the average diffraction $\langle\lambda\rangle$ is equal to zero. The bright soliton solutions exist only for positive average diffraction since we assumed that the nonlinearity $\sigma_\gamma$ is positive. When approaching this limiting value, the spatial extent of solutions grows towards infinity. The second limiting curve in Fig.~\rpict{var2d}(c) indicates the values corresponding to the critical energy $E_{\rm cr}=2\sqrt{\langle\lambda\rangle}$ of the Townes fundamental soliton~\cite{Chiao:1964-479:PRL, Berge:1998-259:PRP, Moll:2003-203902:PRL} for the average diffraction $\langle\lambda\rangle$. Stable solutions exist only for values of $E>E_{\rm cr}$.

For a fixed value of the structural parameter $B$, there exists a family of soliton solutions. These solutions can be characterized by the pules energy $E$, the phase accumulated over one period $\Delta \phi$ (this parameter has a similar meaning to the propagation constant for solitons in homogeneous structures), and the diffraction map strength $S$.
We define the values of $S$ analogous to dispersion map strength in structures with varying temporal dispersion~\cite{Berntson:1998-900:OL, Malomed:2001-1243:JOSB}:
\begin{equation} \leqt{S}
    S=\int_0^L |\lambda(z)| \frac{ dz}{W_{\rm FWHM}^2} ,
\end{equation}
where $W_{\rm FWHM}$ is the full width at half maximum of the wave packet at the point where it is the narrowest along the $x$ direction.
Typical dependencies of $E$ and $S$ on $\Delta \phi$ are shown in Fig.~\rpict{var2d}(d).
The value of $E$ at $\Delta \phi \rightarrow 0$ is equal to the critical energy of the Townes soliton $E_{\rm cr}$.
We observe that the energy is a monotonously growing function of $\Delta \phi$, which indicates that the Vakhitov-Kolokolov criterion~\cite{Vakhitov:1973-783:RQE}
for soliton stability is fulfilled. However this criterion is not sufficient to predict stability for periodically modulated structures, and we find that the solutions become unstable for high values of the energy $E$ and diffraction map $S$.

\pict[1.0]{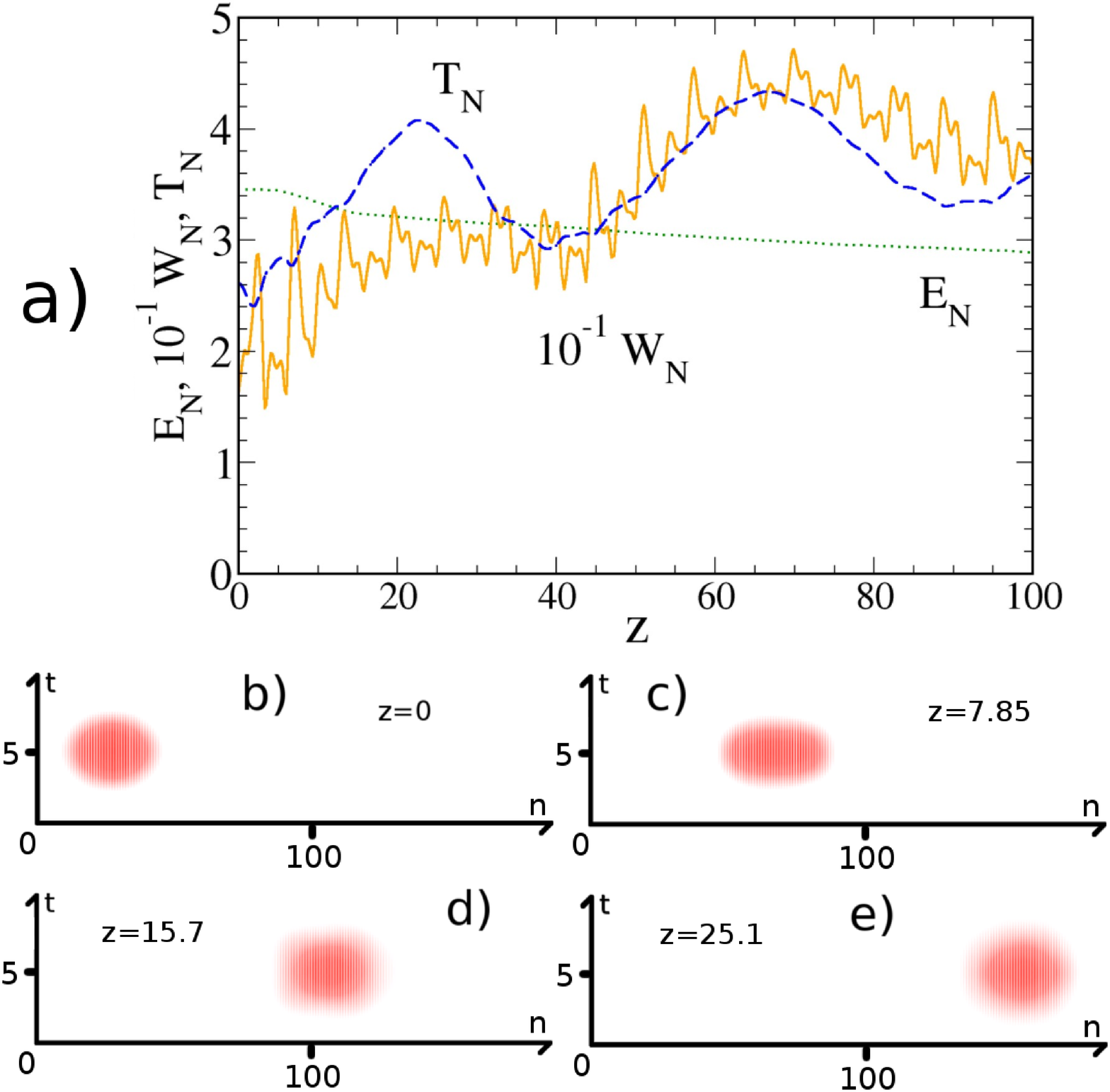}{stable2d}{
Stable two-dimensional light bullet moving across the array. Parameters values are $B=2.1$, $C=100$, $D=\gamma=1$,
and the initial Gaussian shape corresponds to the variational solutions with $E=1.1$, $W(0)=0.68$, $T(0)=1.55$, $b(0)=0$, $T'(0)=0$, multiplied by a spatial chirp function $\exp(i k n)$ with $k = \pi / 20$.
(a) The spatial and temporal widths $W_{\rm N}$ and $T_{\rm N}$, and the total energy $E_{\rm N}$ of the propagating pulse found by the numerical
integration of the discrete propagation equation~\reqt{PE}.
(b)-(e) Snapshots of the spatiotemporal pulse shape at several propagation distances.
}

\subsection{Mobile bullets: full numerical modeling} \lsect{2d-numerics}

We note that the equation~\reqt{PE_red} admits solutions moving in both spatial and temporal dimensions. For example, any solution $u(x,t,z)$ can be multiplied
by a linear phase which gives a new solution $u(x-x_2) \exp[ikx + i \phi_2(z)]$, where
\begin{equation} \leqt{move}
    \frac{d x_2}{d z} = k \lambda(z).
\end{equation}
Hence, pulses can move along $x$-direction during the propagation, and this movement gives rise to an
averaged soliton drift if the mean diffraction $\langle\lambda\rangle$ is nonzero.
However, the original discrete-continuous Eq.~\reqt{PE} does not have such symmetry, and soliton motion can be suppressed due to the Peierls-Nabarro potential associated with the underlying discretness~\cite{Kivshar:1993-3077:PRE, Aceves:1996-1172:PRE, Morandotti:1999-2726:PRL}. Therefore, it is essential to confirm the soliton mobility through direct numerical simulations of Eq.~\reqt{PE}.

In Fig.~\rpict{stable2d} we present an example of a stable light bullet which moves across the array according to numerical simulations of the full discrete-continuous propagation equation~\reqt{PE}.
Here, the initial condition was taken from the periodic solution of the variational equations~\reqt{VA_2D} multiplied by a spatial chirp function $\exp(i k n)$
with $k = \pi / 20$,
which corresponds to the initial beam tilt. As depicted in Fig.~\rpict{stable2d}(a), the spatial
and temporal widths of the soliton oscillate during the propagation in an irregular pattern, but dynamical stability is achieved over long propagation distances.
The widths in the numerical simulations are calculated according to the formula
\begin{equation} \leqt{WN}
    W_{\rm N}=3\langle\, |n - \langle n \rangle|\, \rangle,
\end{equation}
where $\langle n \rangle = \sum_n \int n |a_n(t,z)|^2 dt / \sum_n \int |a_n(t,z)|^2 dt$.
It gives approximately the full-width at half-maximum (FWHM) value for a Gaussian pulse. The numerically calculated energy is defined as
\begin{equation} \leqt{EN}
    E_{\rm N}= d_x \sum_n \int |a_n(t,z)|^2 dt.
\end{equation}
Moving solitons in lattices are known to exhibit radiative losses~\cite{Yulin:2003-260402:PRL}. In our simulations, we observe that the energy of the light bullet slowly decreases mainly due to radiation in the waveguide sections with the highest curvature. This radiation is more pronounced for tighter bending
[smaller values of $C$ corresponding to smaller bending period $L_d$ acording to Eq.~\reqt{widthCondition}]. In Fig.~\rpict{stable2d}(b)-(e) the spatiotemporal profiles of the light bullet are shown for several values of the propagation distance $z$.
Here, the solution is shown in the reference frame moving along the periodic translation with zero mean shift resulting from the waveguide bending.
The stable evolution presented in Fig.~\rpict{stable2d} corresponds to typical physical parameters for optical pulses
$W_d \simeq 100\,{\rm \mu m }$ and $T_d \simeq 50\, {\rm fs}$ in photonic lattice with the physical characteristics of $d_d = 5\, {\rm \mu m}$, $L_d = 10\, {\rm cm}$, $C_d \simeq 6.3\, {\rm mm^{-1}}$, 
and $D_d = 175\, {\rm fs^2\;cm^{-1}}$.

For comparison, in Fig.~\rpict{unstable2d} we show a solution for the same values of the parameters as in Fig.~\rpict{stable2d} except for a lower pulse energy, corresponding to the unstable region in Fig.~\rpict{var2d}(c).
In this case, the pulse quickly spreads out in both space and time.

\pict[1.0]{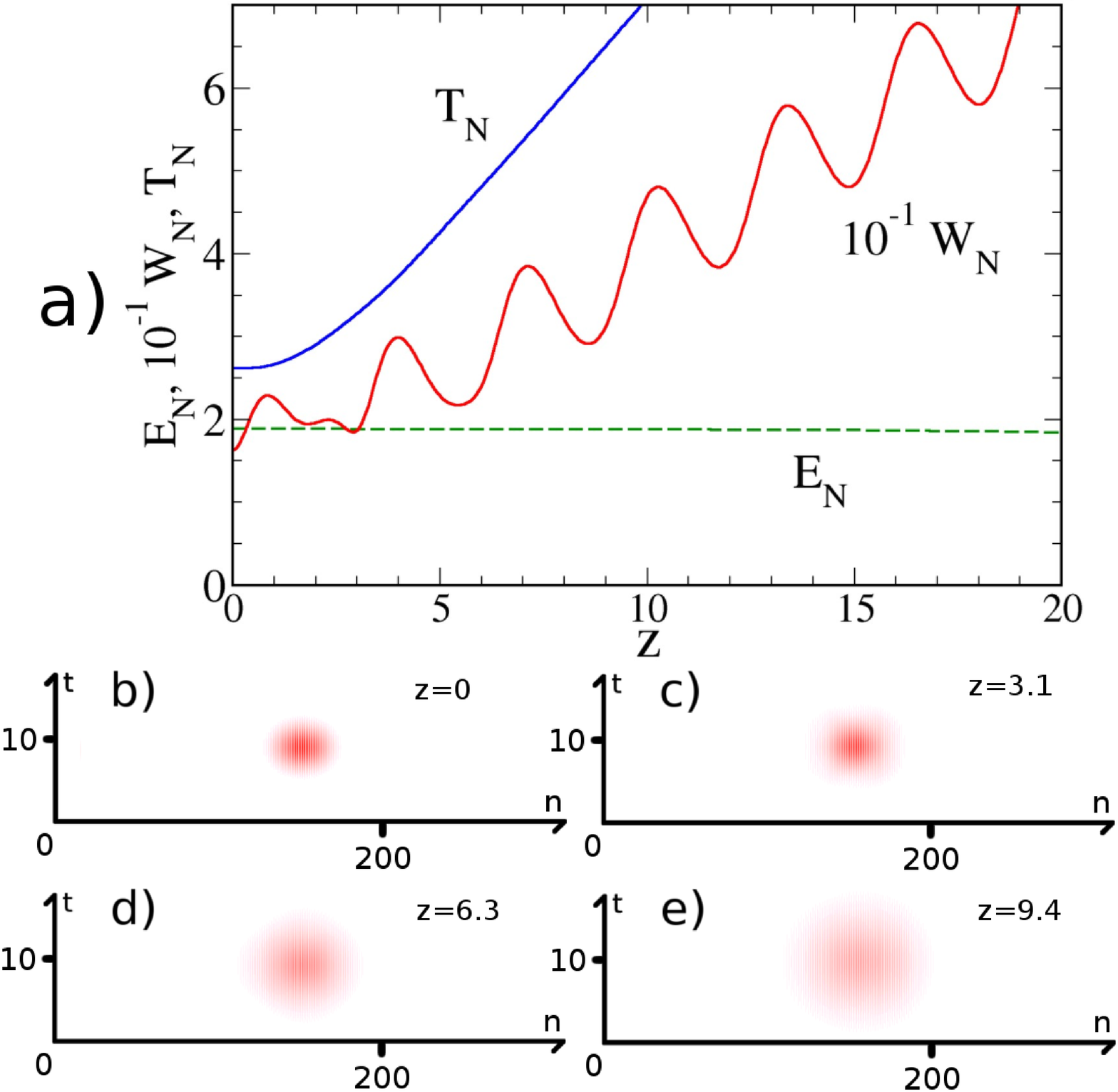}{unstable2d}{
An example of an unstable solution. Parameters are the same as in Fig.~\rpict{stable2d} except $E=0.6$.
}

\section{Three-dimensional light bullets} \lsect{3d}

We now consider the possibility to stabilize three dimensional light bullets, while preserving their mobility. This is a more challenging problem, since two-dimensional light bullets in media with cubic nonlinearity exhibit critical collapse, and even weak perturbations can be sufficient to prevent the collapse effect~\cite{Berge:1998-259:PRP}. On the other hand, the three-dimensional light bullets can exhibit super-critical collapse, which can only be prevented through strong changes in the physical system~\cite{Berge:1998-259:PRP}.

We consider the existence and dynamics of three-dimensional light bullets in the two-dimensional periodically curved square lattice of optical waveguides, as schematically shown in Fig.~\rpict{var3d}(a). For the square lattice, the discrete-continuous propagation equation has the same form as Eq.~\reqt{PE} with the additional coupling along the $y$ axis
\begin{align}
    i \frac{\partial a_{n,m}}{\partial z} &
    + C \left(\ee^{-i p_x(z)} a_{n+1,m} + \ee^{i p_x(z)} a_{n-1,m}\right) + \nonumber \\
    &+ C \left(\ee^{-i p_y(z)} a_{n,m+1} + \ee^{i p_y(z)} a_{n,m-1}\right) + \\
    &+ \frac{D}{2} \frac{\partial^2 a_{n,m}}{\partial t^2} + \gamma |a_{n,m}|^2 a_{n,m}
    =0 \leqt{PE_3D}\,, \nonumber
\end{align}
where $y = y_d / y_s$ is the normalized coordinate, and we assume the same scaling of the transverse spatial dimensions, $y_s = x_s$. Then, the coupling phase coefficients are defined similar to Eq.~\reqt{PEphase},
\begin{equation}  \leqt{PEphase3d}
    p_x(z) = \frac{2 \pi n_0 d x_s^2}{\lambda_d z_s} x^{\prime}_0, \,
    p_y(z) = \frac{2 \pi n_0 d x_s^2}{\lambda_d z_s} y^{\prime}_0,
\end{equation}
where $x_0(z)$ and $y_0(z)$ define the transverse bending profile.

We further consider the helical waveguide bending profile [see Fig.~\rpict{var3d}(a)],
\begin{equation} \leqt{xy0Helical}
    x_0(z) = x_{0m} \cos ( 2 \pi z / L ),\,
    y_0(z) = y_{0m} \sin ( 2 \pi z / L ),
\end{equation}
where $x_{0m} = y_{0m}$ is the bending amplitude (normalized to $x_s$) and $L$ is the normalized bending period.

Following the approach described above in Sec.~\rsect{2d-continuous}, we consider wavepackets that are slowly varying in space and describe their profiles with a continuous function,
$a_{n,m}(t,z) = \tilde{u}(x=n d,y = m d, t,z)$. Then, we apply the scaling coefficients according to Eq.~\reqt{scalingCoeff}, and derive the approximate continuous equation [cf. Eq.~\reqt{PE_red}]:
\begin{align}
    i \frac{\partial u}{\partial z}
    + \frac{\lambda_x(z)}{2}  \frac{\partial^2 u}{\partial x^2}
    + \frac{\lambda_y(z)}{2}  \frac{\partial^2 u}{\partial y^2}
    +\frac{\sigma_D}{2} \frac{\partial^2 u}{\partial t^2}
    + \sigma_\gamma |u|^2 u =0,
\end{align}
where $\lambda_x(z)$ and $\lambda_y(z)$ the normalized diffraction coefficients corresponding to the two transverse directions:
\begin{equation} \leqt{lambda3d}
    \begin{array}{l} {\displaystyle
        \lambda_x(z) = \cos (B \sin z),\,
    } \\*[9pt] {\displaystyle
        \lambda_y(z) = \cos (B \cos z),\,
    } \\*[9pt] {\displaystyle
        B = - x_{0m} \frac{4 \pi^2 n_0 d x_s^2}{\lambda_d z_s}.
    } \end{array}
\end{equation}

Next, we perform the variational analysis using the three dimensional Gaussian Ansatz in the following form:
\begin{eqnarray}
    u &=&A(z)\exp \left\{ \mathrm{i}\phi (z)-\frac{1}{2}\left[
    \frac{x^{2}}{W^{2}(z)}+\frac{y^{2}}{V^{2}(z)}+\frac{t
    ^{2}}{T^{2}(z)}\right] \right. +
    \nonumber \\
    &&+\left. \frac{\mathrm{i}}{2}\left[ b(z)\,x^{2} + c(z)\,y^{2} +\beta (z)\,t
    ^{2}\right] \right\} ,  \leqt{ansatz3d}
\end{eqnarray}
Following similar procedure as we used in the two-dimensional case [Sec.~\rsect{2d-variational}], we arrive at the set of variational equations
\begin{eqnarray}
    W^{\prime }&=&\lambda_1 b W,  \nonumber  \\
    V^{\prime }&=&\lambda_2 c V,   \nonumber \\
    b^{\prime } &=&\frac{\lambda_1}{W^{4}}-\lambda_1 b^2 -\frac{E}{2^{3/2}W^{3}VT},  \leqt{VA_3D} \\
    c^{\prime } &=&\frac{\lambda_2}{V^{4}}-\lambda_2 c^2 -\frac{E}{2^{3/2}WV^{3}T},  \nonumber \\
    T^{\prime \prime } &=&\frac{1}{T^{3}}-\frac{E}{2^{3/2}WVT^{2}},  \nonumber
\end{eqnarray}
where $E\equiv A^{2}WVT$.

\pict[1.0]{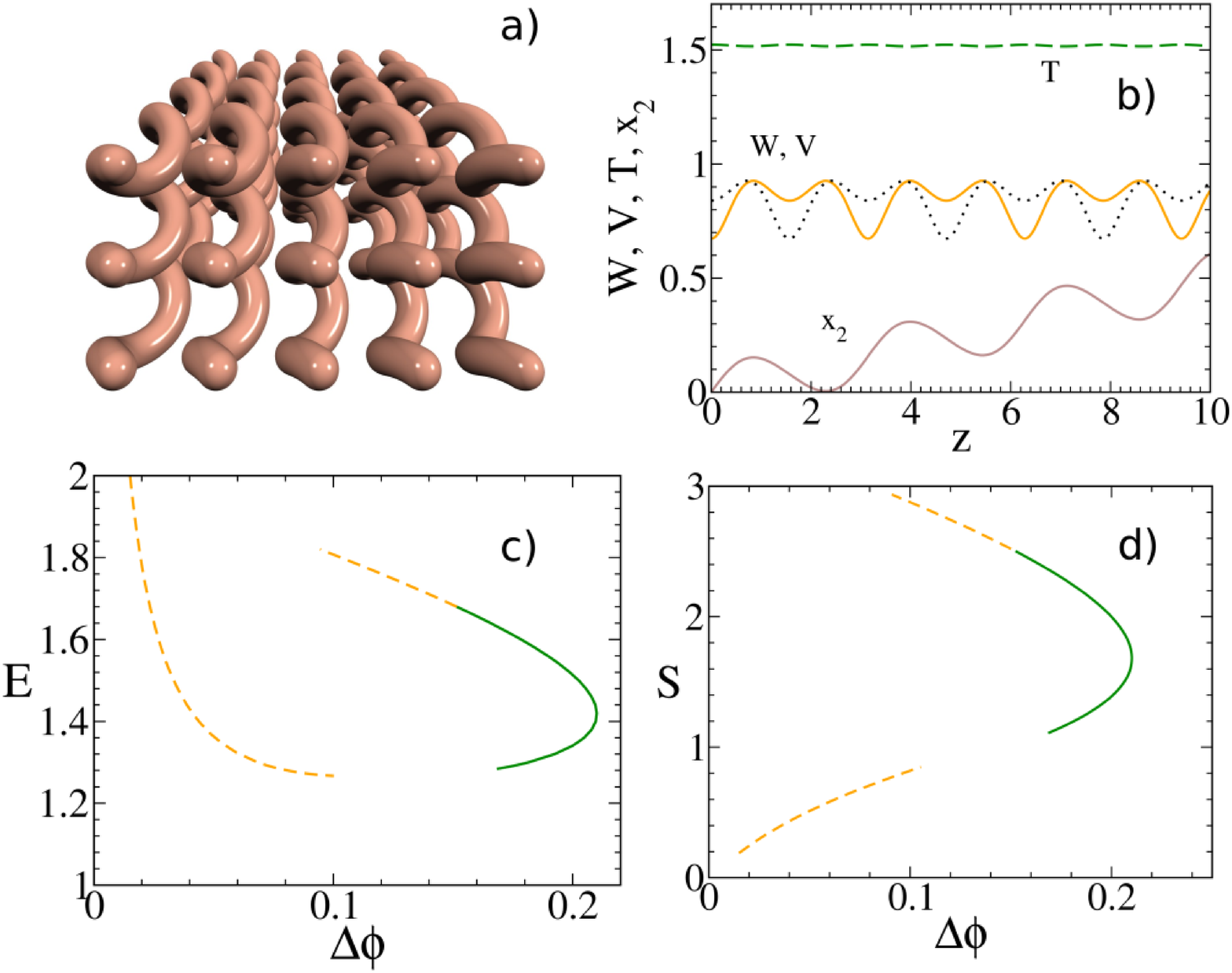}{var3d}{
(a) Sketch of helical  waveguide array in three dimensions. (b) The dependence between the soliton  energy $E$ and the phase increment $\Delta \phi$
for
one bending period. Solid line corresponds to the stable solutions, dashed line corresponds to the unstable solutions. (c) Stable three-dimensional light bullet moving across the helical  array.
The transverse spatial widths $W$ (solid), $V$ (dotted) and the
temporal width $T$ (dashed) of the light bullet are shown as functions of the propagation distance $z$. The solid line
below shows the position of the moving pulse according to Eq.~\reqt{move} for $k_x=0.3$.
}

We perform numerical analysis to identify periodic solutions of the variational equations~\reqt{VA_3D}. Such solutions correspond to three-dimensional light bullets, where the light-bullet 
extension along all dimensions ($x,y,t$) is restored after each bending period. We present an example of such solution in Fig.~\rpict{var3d}(b), which illustrates a stable and mobile three-dimensional light bullet.
 Indeed, the light bullet spatial and temporal localization widths are restored to the same values after each bending period.
The solid line at the bottom shows the transverse position $x_2$
of the pulse center as a function of the propagation distance $z$ for a finite
initial beam tilt. The light bullet wobbles in the transverse direction during the propagation, and it is clear that on average the pulse moves across the array.

We summarize the existence and stability properties of three-dimensional light bullets in Figs.~\rpict{var3d}(c,d). We show the dependence between the energy $E$ and the diffraction map strength $S$
on the phase increment $\Delta \phi$ for one bending period. In comparison with the two-dimensional case [Fig.~\rpict{var2d}(d)], this dependence is much more complex, forming a loop-like structure. Interestingly, stable solutions, depicted with a solid line, are present for both positive and negative curve slopes of the $E(\Delta \phi)$ dependence, which indicates that the Vakhitov-Kolokolov criterion~\cite{Vakhitov:1973-783:RQE} 
 is not applicable here.

\section{Matter-wave solitons} \lsect{BEC}

Finally, we note that a very related problem in the field of Bose-Einstein condensates is the stabilization of multidimensional matter-wave solitons, which dynamics can be described by the discrete Gross–Pitaevskii equation which is mathematically equivalent to Eq.~\reqt{PE}.
Several methods of stabilization have been considered to date. Discreteness can be introduced by
applying deep optical lattices, with the cost of reducing the soliton mobility to one dimension~\cite{Baizakov:2004-53613:PRA}. On the other hand, shallow optical lattices can
be used to manipulate the effective mass of the mater waves~\cite{Eiermann:2003-60402:PRL, Fallani:2003-240405:PRL, Lignier:2007-220403:PRL}, allowing for stabilization of dispersion-managed solitons in two dimensions~\cite{Abdullaev:2003-66605:PRE}.
Stabilization of subdiffractive solitons with almost zero dispersion in lattices in two dimensions was also proposed~\cite{Staliunas:2007-11604:PRA}.
Another approach consists of modulation
of the nonlinear coupling using Feshbach resonances. In this context, stable two dimensional solitons were predicted~\cite{Abdullaev:2003-13605:PRA, Saito:2003-40403:PRL, Itin:2006-33613:PRA}, however
stabilization in the full three-dimensional case requires an additional potential~\cite{Montesinos:2004-193:PD, Trippenbach:2005-8:EPL, Matuszewski:2005-50403:PRL},  feedback control of the
nonlinearity~\cite{Saito:2006-23602:PRA} or inclusion of inelastic losses~\cite{Saito:2004-53610:PRA}. Nonlocal dipolar interactions were also shown to provide stabilization
of quasi two dimensional BEC solitons~\cite{Pedri:2005-200404:PRL, Tikhonenkov:2008-90406:PRL}.

Our method can provide an alternative solution to the stabilization of matter-wave solitons. We estimate that
in the case of a Bose-Einstein condensate of $^7$Li atoms in the quasi two dimensional setting with the axial trapping frequency $\omega_{\perp}=2\pi\times 300\,$Hz,
the lattice constant has the value $d=\lambda/2=0.4\,\mu$m, the width along the lattice axis
$W_x=6.5\,\mu$m, transverse width $W_y=20.9\,\mu$m and the number of atoms $N=4000$, where we assumed that the scattering length is tuned to
$a_{\rm s}=-5 a_0$ using a Feschbach resonance \cite{Pollack:2009-90402:PRL}. The time period of the lattice modulation is equal to $\tau=45\,$ms and the total time of
evolution in Fig.~\rpict{stable2d}(a) is $0.7\,$s.

\section{Conclusions} \lsect{conclusions}

We have demonstrated that arrays of curved waveguides can be used to stabilize two- and three-dimensional light bullets against the collapse in media with Kerr-type nonlinearity. Most remarkably, the bullets can freely move across the periodic structure, unlike the previously considered discrete light bullets which were trapped to a specific lattice site. These results suggest new possibilities for flexible spatio-temporal manipulation of optical pulses, and they are also relevant for the stabilization of multidimensional matter-wave solitons.

This work was supported by the Australian Research Council.

\end{sloppy}
\end{document}